# Electron spectroscopic studies of nanowires formed by (GaMn)As growth on GaAs(111)B

J. Adell[1], I. Ulfat[1,2], J. Sadowski[3,4], L. Ilver[1], and J. Kanski[1]

[1] *Department of Applied Physics, Chalmers University of Technology, Göteborg, Sweden*

[2] *Department of Physics, University of Karachi, Karachi, Pakistan*

[3] *Institute of Physics, Polish Academy of Sciences, Warszawa, Poland*

4. *MAX-Lab, Lund University, 221 00 Lund, Sweden*

(Date: 1. July. 2009)

Valence band photoemission with photon energies around the Mn2p excitation threshold has been used to study the development of nanowires catalyzed by MnAs particles. A gradual change in the spectra with increasing nanowire length is observed, such that the resonant photoemission eventually dominates over the Auger decay channel. The change is ascribed to dilution of Mn, showing that Mn is transferred from the MnAs particles into the nanowires.

**Introduction**

The interest in semiconductor based self-assembled nano-structures has increased rapidly during the last years [1, 2]. In particular nanowires (NW) or -whiskers are considered to have great promise in a wide range of applications [1, 3, 4]. The first description of NW growth, the so called vapor-liquid-solid (VLS) mechanism was based on the existence of eutectic droplets, from which one of the NW components is expelled due to supersaturation [5]. Since then, growth techniques have evolved and recently it has been found that NW growth can also be catalyzed by solid particles [6, 7]. In view of the potentially wide range of applications, it is obvious that the physical mechanisms of self-assembled NWs growth must be well understood and eventually controlled. Concurrently there has been a renewed interest in magnetic semiconductors during recent years, largely motivated by the demand for functional materials for future spin-based electronics. In this context diluted magnetic semiconductors such as (GaMn)As have been particularly attractive, as they are compatible with materials used in devices of present days information technology. (GaMn)As has also already been demonstrated to be suitable for fabrication of spintronic devices, for example spin lasers [8] or tunneling magnetoresistance structures [9]. Attempts to combine these two fields of research and produce one-dimensional spintronic devices have recently been explored on for example ZnO [10] by post growth Mn implantation.

In the present study NWs were formed during growth of (GaMn)As at a temperature above that used for growth of diluted (GaMn)As. Under these conditions particles are segregated on the surface and catalyze nanowire growth [6]. The key issue addressed here is the incorporation of Mn in the nanowire, i.e. whether Mn is diluted, or remains in the form of MnAs. Considering the poor solubility of Mn in GaAs and the fact that the segregated state is more stable than the mixed phase, dilution of Mn from the MnAs in the nanowire tip is not expected.

**Experiments**

The experiments were carried out at BL 1011 at the Swedish synchrotron radiation laboratory MAX-lab, where a system for molecular beam epitaxy (MBE) is available at one of the beamlines for sample production. The samples were grown on epiready GaAs(111)B wafers, which were indium glued to Mo blocks. A small ion-pumped transport chamber is used for transfer to other beamlines, such that the samples are always kept in UHV after growth. The preparation of the NW samples has been described earlier [6]. Briefly, after standard surface preparation including oxide desorption and buffer layer growth, the substrate temperature ($T_S$), monitored with an IR pyrometer, was decreased to 350 °C and (GaMn)As growth was started with the Mn source set at a temperature corresponding to growth of diluted (GaMn)As with 1% Mn. Since the growth temperature exceeded the maximum temperature for 2D growth, a transition from 2D to 3D growth was observed in RHEED indicating formation of MnAs particles. AFM images of a surface at this stage show islands with 300-500 Å diameter and 20 – 50 Å height [6]. With continued growth under these conditions and using GaAs(111)B substrates, the RHEED pattern developed into that expected for NWs extending along the surface normal, with sharp spots along the NW direction and streaks in the orthogonal direction [6].

Data from two kinds of samples will be discussed here, obtained on NWs grown with and without Mn supply after the formation of MnAs particles. In the first set, with Mn supply, NW lengths ranging from 200 nm to 1600 nm were obtained by varying the growth period from 15 min to 120 min. In the second set, without Mn supply, the growth lasted for 75 min. Figure 1 shows SEM images from two samples grown under the different conditions. In addition to the NW samples we also prepared two reference samples, a 500 Å thick non-segregated (GaMn)As film with 5 % Mn, and an approximately 300 Å thick MnAs film.

The samples were initially characterized by X-ray absorption (XAS) around the Mn2p threshold. The spectra were recorded in the total yield mode with exit slit of the Zeiss monochromator set to give an energy resolution of 150 meV. The XAS spectra were used to select photon energies for the subsequent photoemission experiments. The photoemission data were recorded with a Scienta SES200 electron energy analyzer, operated at a constant pass energy of 100 eV. The combined energy resolution was around 400 meV.

**Results and discussion**

From Fig. 1 we see a clear distinction between the two types of wires. The growth rate of the second set of wires is greatly reduced relative to when Mn is supplied during the wire growth, and it is also observed that without Mn supply the wires grow only little faster than the surrounding flat surface such that each wire protrudes from a hollow region in the surface. These observations show that Mn supplied during growth stimulates surface diffusion, such that the uptake surface area for the nanowires is largely extended by the presence of Mn.

As mentioned above, the key issue of the present study is the distribution of Mn in NWs formed under MnAs particles. In a previous report [6] an EDX line scan of the elemental distribution of Mn, Ga and As in an individual wire grown under the similar conditions showed that most of the Mn remained at the NW tip, probably in the form of MnAs. On the other hand one could note that the shape of NWs formed in this way is clearly tapered, significantly narrowing towards the tip, which means that either the size of the catalyzing particle is gradually decreased during the NW growth, or due to the low growth temperature, the Ga adatoms diffusing along the NW walls towards the tips are incorporated into the NWs before reaching the catalyzing nanoparticle [11]. In fact this mechanism can be favored by the relatively low growth temperature. Conversely, the low diameters of the NW tips (below 10 nm) indicate that the diameters of MnAs nanoparticles indeed decrease during the NW growth, since the initial diameters of MnAs nanoclusters before the NW growth start are much higher [6]. This suggest that Mn is either consumed during the growth and distributed along the wire, or concentrated in a more Mn rich compound (e.g. $Mn_2As$). Although the latter scenario would be consistent with the line scan data, the photoemission results shown below clearly favor the first alternative.

To demonstrate the ability of photoemission to distinguish between the two situations we first show results from our two reference samples, containing Mn in a concentrated (MnAs) and in a diluted form (GaMn)As, see Fig. 2. In XAS we can clearly see that the two samples differ from each other. The XAS from MnAs, is clearly much broader than that from (GaMn)As. This broadening can be directly ascribed to band structure formation in the binary compound. In the diluted system, on the other hand, the Mn 3d states are more atomic-like, subject only to crystal field splitting. This is reflected by the narrower spectra, in which the fine structure is due to multiplets of the Mn $d^5$ state in tetrahedral coordination [12]. In the valence band spectra recorded around the Mn 2p excitation threshold the localization of the intermediate $2p^5 3d^{N+1}$ state in the diluted case is manifested by the appearance of resonant photoemission and absence of the incoherent Auger decay. The latter is recognized in the spectra by a structure appearing at constant kinetic energy. In a series of spectra of spectra represented on a binding energy scale this structure therefore shifts with the photon energy (indicated by vertical bars in Fig. 2). This is clearly the situation for the MnAs data. In contrast, the set of spectra from (GaMn)As shows intensity modulations of structures appearing at constant binding energies, which is the signum of resonant photoemission, and no structures are seen at the expected positions of the Auger peak.

We turn next to the NWs. Spectra corresponding to those discussed above were recorded after four growth periods, 15, 30, 60, and 120 min. Since the data from the 15 min and 120 min samples were qualitatively similar to those from the 30 min and 60 min ones, respectively, only the latter are shown here. The X-ray absorption spectra from the two samples are displayed in Fig. 3 a-b). The development of the narrow peak around 640 eV and the more pronounced structures around 650 eV reflect a partial change from MnAs-like to (GaMn)As-like appearance with increasing deposition time (i.e. with increasing nanowire length).

Fig. 3 c-f) shows the corresponding valence band photoemission spectra. Since the samples were obtained under continued supply of Mn, an immediate point of concern is the fact that Mn is present not only in the nanowire tips, but also on the flat part of he sample. Although the NWs appear to be very dense (Fig. 1), one cannot exclude appreciable emission from the surface. To estimate this, we have recorded spectra in normal emission, where the projected NW area is at minimum, and at a large emission angle (i.e. nearly parallel with the surface), in which case the underlying surface should be essentially invisible from the analyzer. For the short NWs a clear angular dependence is seen, particularly in spectra D and E, obtained just above the resonance peak: the broad Auger peak dominates at large angle of emission, showing that the wires are still terminated by MnAs particles. The relatively small Fermi edge step indicates that the MnAs is in ZB structure rather than hexagonal [13]. However, for the longer wires it is clear that the resonant component becomes stronger, and no angular dependence is seen. Also the Fermi edge emission becomes less pronounced. These observations indicate a more dispersed distribution of Mn in the long NWs.

Photoemission spectra from the second type of NWs, grown without Mn supply after the nucleation of MnAs, are shown in Fig. 4. For these samples one can safely exclude any surface contribution to the Mn signal after NW growth, since the surface is covered with GaAs. As already noted, the development of the NWs without Mn supply is strongly hampered, which clearly demonstrates the surfactant role of Mn for surface diffusion of Ga (and possibly Mn). Despite the lower tendency to NW growth, data from these wires provide information that clearly supports the conclusions reached above.  Fig. 4a shows the valence band spectra of the MnAs particles, obtained before the NW development. As previously, the Auger decay channel clearly dominates. However, at the stage of the SEM image in Fig. 1, i.e. after 75 min. deposition of GaAs on the same surface, we observe that the Auger emission is completely quenched and is replaced by resonant photoemission, Fig. 4b. This must be ascribed to dilution of the Mn initially located in the MnAs particles.

**Conclusion**

The data provide new information about the properties of NWs formed by (GaMn)As and catalyzed by MnAs particles. The short wires are terminated by metallic MnAs particles, which act as the catalysts for nanowire growth. From the qualitative modifications of XAS and photoemission found during NW growth, we can conclude that the Mn becomes diluted much like that in (Ga,Mn)As films. Similar behavior is observed for NWs formed with and without Mn supply during the growth, but it is clear that the presence of Mn on the surface strongly promotes NW growth.

**Acknowledgement**

The present work is part of a project supported by the Swedish Research Council (VR). One of the authors (J.S.) acknowledges the financial support from the Ministry of Science and Higher Education (Poland) through grant N N202 126035.

**Figure captions**

Figure 1: a) SEM image of NWs formed by deposition of (GaMn)As on GaAs(111)B after 60 min growth at a deposition rate corresponding to a planar growth rate of 200 nm/h. The wires are around 700 nm long. b) SEM of NWs grown for 75 min without Mn supply.

Figure 2: XAS and PES data from layers of MnAs (in hexagonal phase) and (GaMn)As. The arrows in the XAS graphs indicate energies used in RPES, and the short bars in the PES graphs indicate the expected positions of Auger emission.

Figure 3: a-b) XAS data from NWs. The arrows indicate energies used in RPES. The spectra were recorded with the light incident at 45° relative to the surface, which is at approximately the same projection as the SEM image in Figure 1. c-f) RPES data from nanowire-covered GaAs after 30 min and 60 min growth recorded in near normal emission (left) and at around 85° emission angle (right). Dashed lines show the expected positions of Auger emission.

Figure 4: a) RPES spectra of MnAs clusters formed on the surface after short growth period of (GaMn)As at elevated growth temperature. Resonance is determined from XAS to 642.2 eV and we can clearly see that the Auger channel is dominating similarly to the MnAs reference sample in Fig. 2. b) After 75 min growth of GaAs, NWs are formed catalyzed by the MnAs particles. Unlike in a), we now see a resonant enhancement to the Mn 3d states.

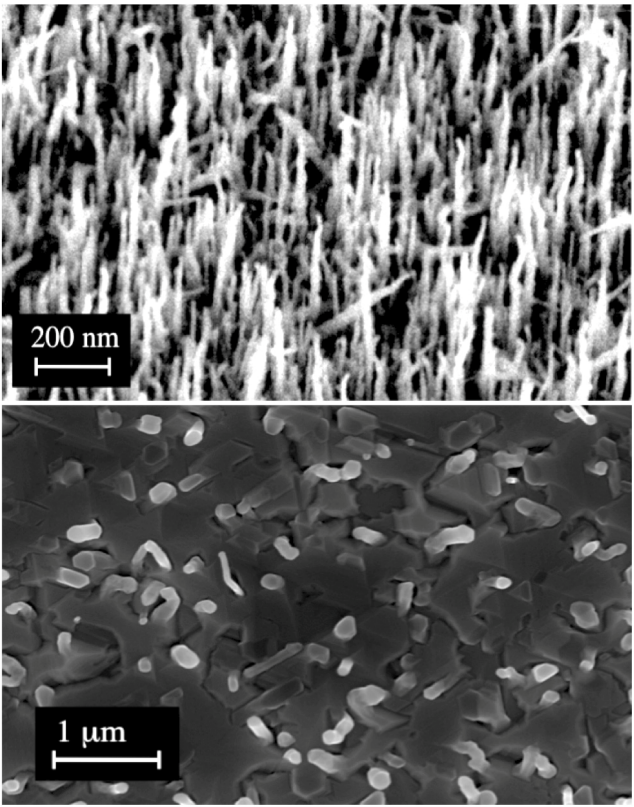

**Figure 1**

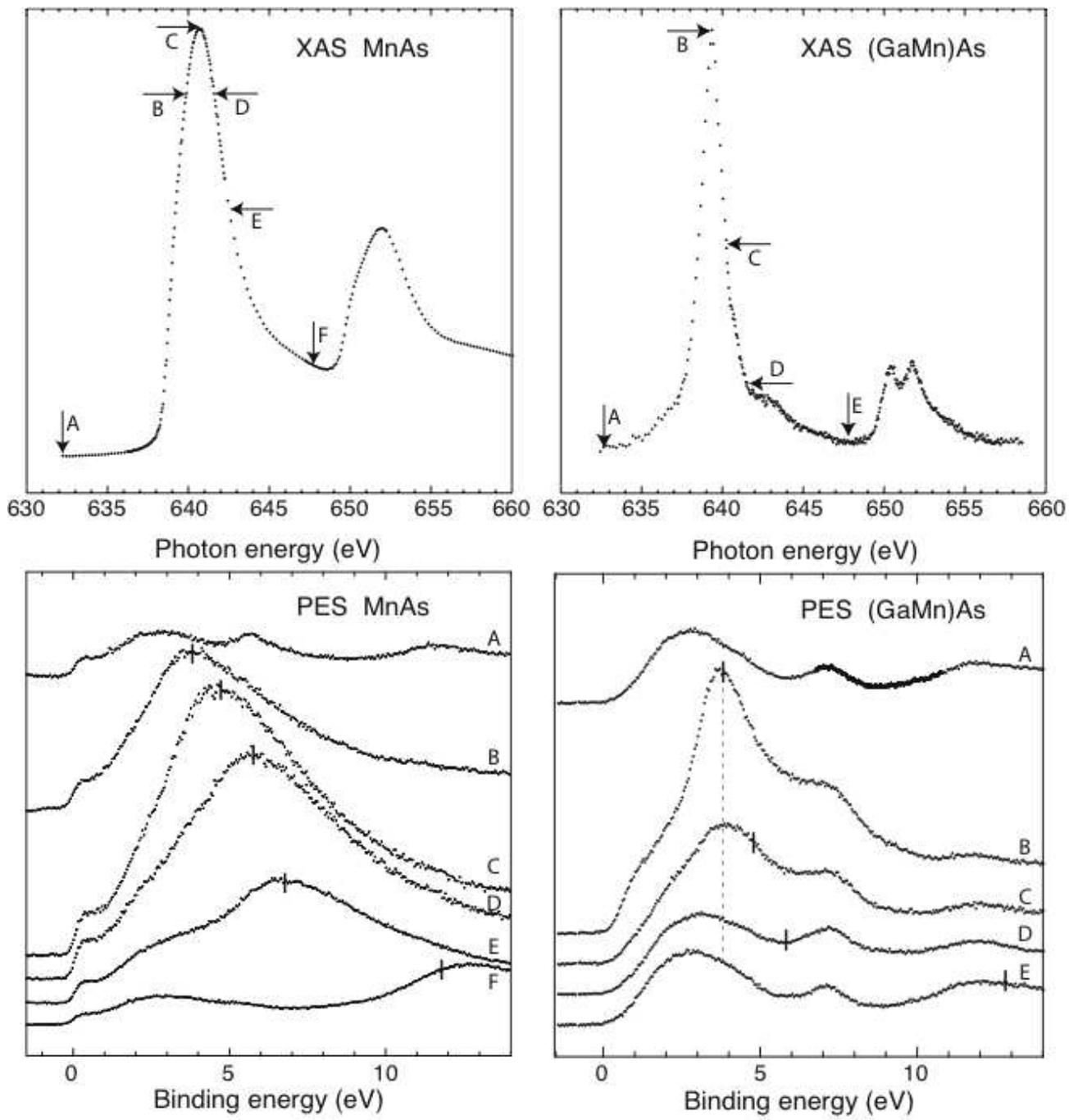

**Figure 2**

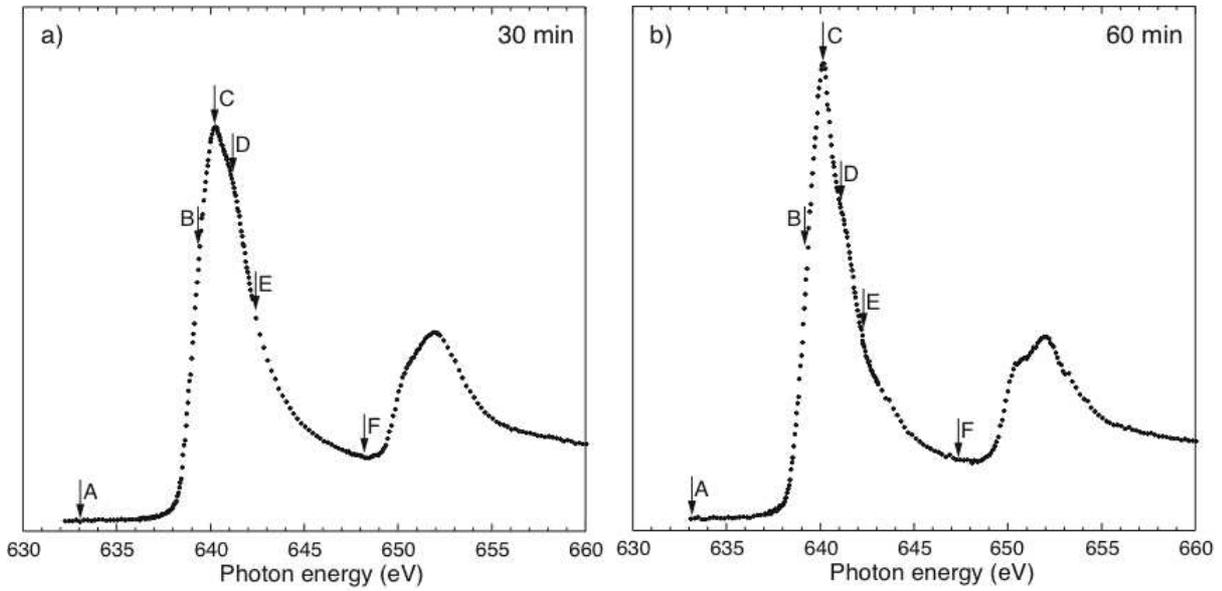
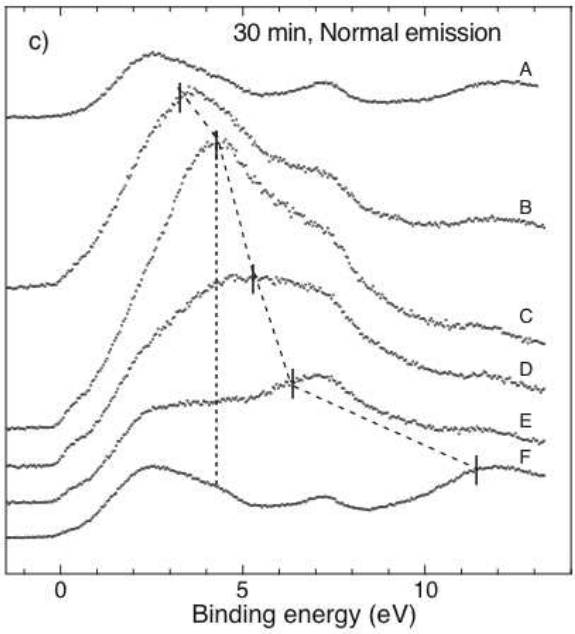
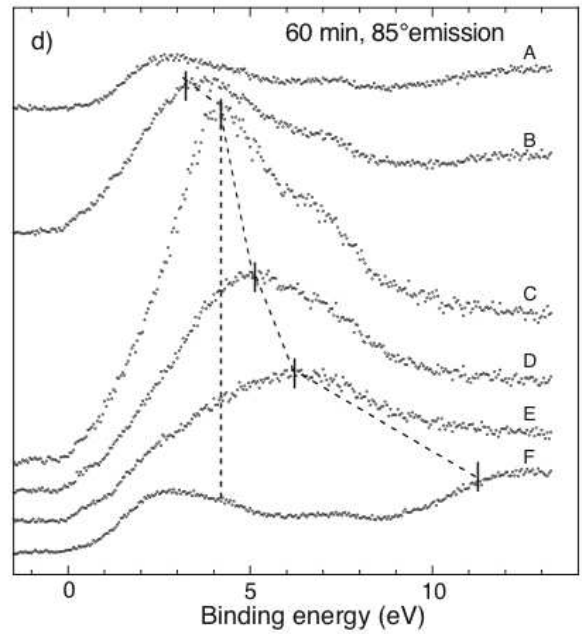
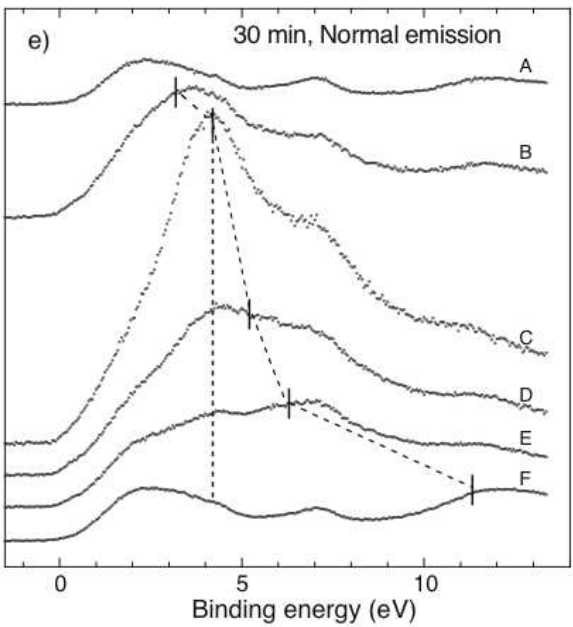
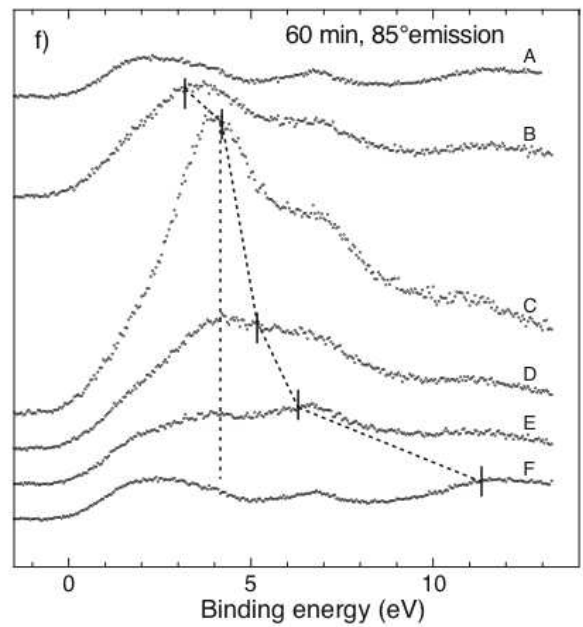

**Figure 3**

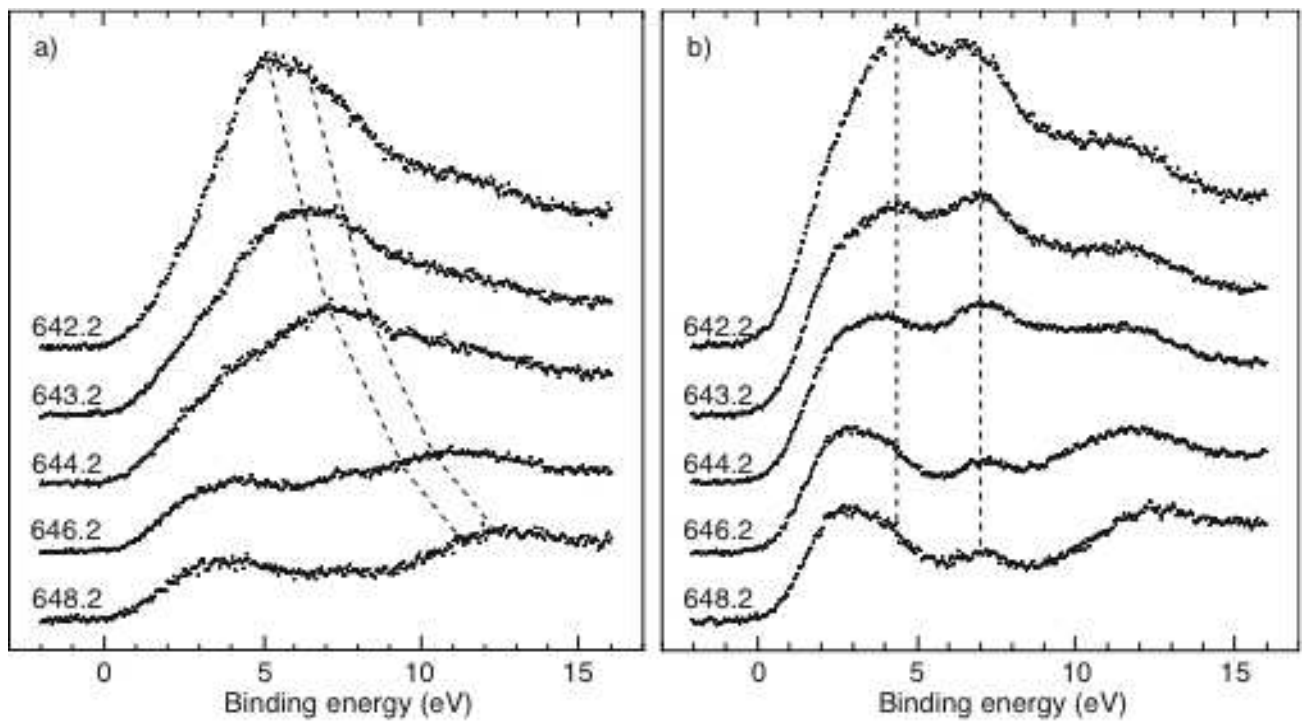

**Figure 4**